\documentclass[fleqn,10pt]{wlscirep}
\usepackage[utf8]{inputenc}
\usepackage[T1]{fontenc}
\usepackage{braket}
\usepackage{qcircuit}

\title{A Digital Quantum Algorithm for Jet Clustering in High-Energy Physics}

\author[1,$\dagger$]{Diogo Pires}
\author[2,3,$\dagger\dagger$]{Pedrame Bargassa}
\author[1,4,$\ddagger$]{João Seixas}
\author[1,2,$\ddagger\ddagger$]{Yasser Omar}

\affil[1]{Instituto Superior T\'{e}cnico, Universidade de Lisboa, Portugal}
\affil[2]{Instituto de Telecomunica\c{c}\~oes, Physics of Information and Quantum Technologies Group, Lisbon, Portugal}
\affil[3]{Laboratório de Instrumentação e Física Experimental de Partículas (LIP), Lisbon, Portugal}
\affil[4]{Centro de Física e Engenharia de Materiais Avançados (CeFEMA), Instituto Superior Técnico, Av. Rovisco Pais 1, 1049-001 Lisboa, Portugal}

\affil[$\dagger$]{diogofgpires@tecnico.ulisboa.pt}
\affil[$\ddagger$]{joao.seixas@tecnico.ulisboa.pt}
\affil[$\dagger\dagger$]{bargassa@cern.ch}
\affil[$\ddagger\ddagger$]{yasser.omar@lx.it.pt}

\begin{abstract}

Experimental High-Energy Physics (HEP), especially the Large Hadron Collider (LHC) programme at the European Organization for Nuclear Research (CERN), is one of the most computationally intensive activities in the world. This demand is set to increase significantly with the upcoming High-Luminosity LHC (HL-LHC), and even more in future machines, such as the Future Circular Collider (FCC). As a consequence, event reconstruction, and in particular jet clustering, is bound to become an even more daunting problem, thus challenging present day computing resources. In this work, we present the first digital quantum algorithm to tackle jet clustering, opening the way for digital quantum processors to address this challenging problem.
Furthermore, we show that, at present and future collider energies, our algorithm has comparable, yet generally lower complexity relative to the classical state-of-the-art $k_t$ clustering algorithm.

\end{abstract}

\begin{document}

%TC:ignore
\flushbottom

\maketitle

\thispagestyle{empty}

%TC:endignore

\section{Introduction}

% In a world where big data is inevitably becoming the norm of the everyday technological landscape, computational tasks are bound to become increasingly intense. Following the trend, data processing and analysis in experimental high-energy physics, and in particular at the LHC programme at CERN, is no exception\cite{HLLHC}. Given the small production cross section of the events of interest associated to New Physics (NP), it is necessary to study a large number of interactions, for which an enormous number of collisions per second and per unit area of the beam is needed. Due to the HL-LHC upgrade currently ongoing, the computational resources demand is set to increase drastically, resulting in increased track multiplicities, with the Compact Muon Solenoid (CMS) tracker being expected to reach hits of the order of $\sim10^3$\cite{hitsHLLHC}. As a consequence, event reconstruction, and in particular jet clustering, is bound to become an even more daunting combinatorial problem, thus challenging present day computing resources.

In a world where big data is inevitably becoming the norm of the everyday technological landscape, computational tasks are bound to become increasingly intense. Following the trend, data processing and analysis in experimental high-energy physics, and in particular at the LHC, presents some of the most computationally challenging tasks worldwide. Given the small production cross section of the events of interest associated to New Physics (NP), it is necessary to analyze an enormous number of events, often very complex in structure. As such, the situation regarding computational resources is bound to become drastically more demanding after the HL-LHC upgrade currently under way, with event sizes being expected to increase $\sim 10$ fold\cite{LHC,HLLHC}. Consequently, event reconstruction, and in particular jet clustering, is bound to become an even more daunting combinatorial problem, thus challenging present day computing resources.

\begin{figure}[ht]
\hbox{\hspace{-1em}
    \centering
    \includegraphics[scale = 0.33]{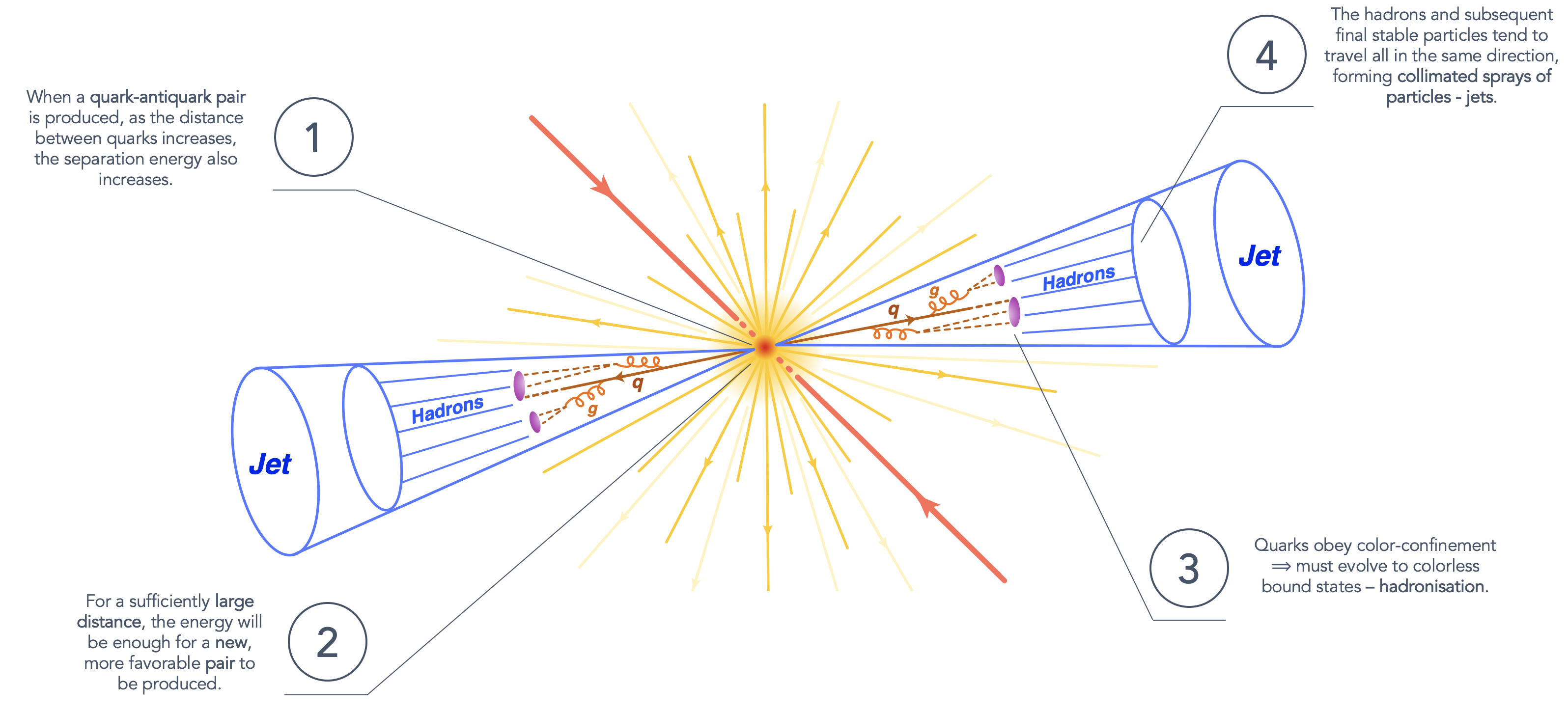}}
    \caption{Example of a Dijet event (\textit{e.g.} from an $e^+e^-$ collision), where a quark-antiquark pair is produced, later giving origin to colorless bound states through hadronisation, and resulting in two back-to-back jets.}
    \label{fig:jet}
\end{figure}

A jet algorithm maps the momenta $\{\Vec{p}_i\}$ of $N$ collimated final-state particles, into the momenta $\{\Vec{j}_k\}$ of $K$ clusters called jets, dependent on the collision conditions and the subsequent particles' distribution, in an approximate attempt to reverse-engineer the quantum mechanical processes of fragmentation and hadronisation as a way of probing the underlying Quantum Chromodynamics (QCD) processes (see Figure \ref{fig:jet}). Recently, quantum algorithmic approaches for jet clustering have been proposed, namely in a quantum annealing formulation\cite{Harrow, Pires}. In this work, we develop a digital quantum algorithm for multijet clustering. Namely, we propose a new, modified, version of the quantum \textit{k-means} algorithm\cite{qkmeans} to address the jet clustering problem. Furthermore, we implement and classically simulate it through the use of IBM's Qiskit package\cite{qiskit}. Finally, we benchmark the performance of our quantum algorithm against the classical state-of-the-art $k_t$ jet clustering algorithm\cite{kt}, namely in terms of its scaling, as well as in terms of its clustering efficiency.

\section{Algorithm}

% Up to three levels of \textbf{subheading} are permitted. Subheadings should not be numbered.

\subsection{Classical \textit{k-means}}

\begin{figure}[ht]
    \centering
    \includegraphics[scale = 0.44]{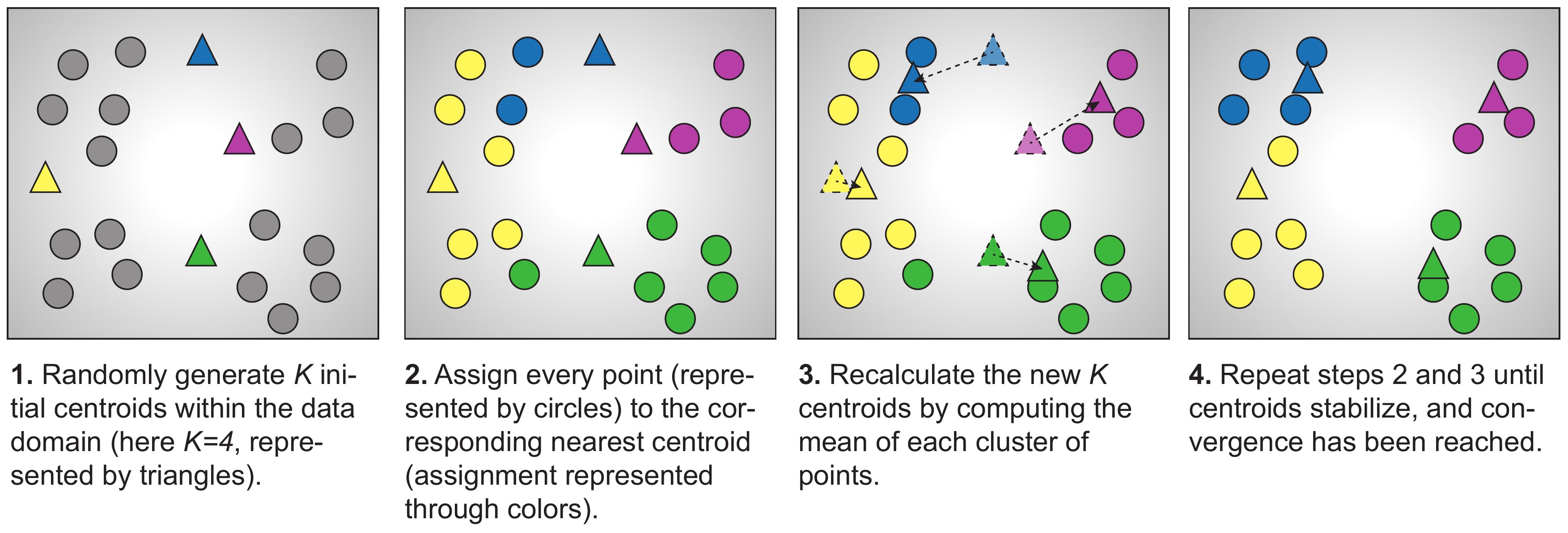}
    \caption{Overview of the procedure relative to the classical \textit{k-means} algorithm.}
    \label{fig:kmeans}
\end{figure}

The classical \textit{k-means} algorithm\cite{kmeans} (see Figure \ref{fig:kmeans}), applied to jet clustering for the first time in 2006\cite{Chekanov} and subsequently in 2012\cite{Thaler_2012} and 2015\cite{Stewart_2015}, receives as input a set of $N$, $D$-dimensional data points and outputs $K$ centroids, calculated through the mean of each group of data points, thus defining $K$ clusters. To be assigned to any particular cluster, a data point needs to be closer to that cluster's centroid than to any other centroid in the data set. In order to successfully converge to the final set of centroids, the algorithm iteratively alternates between assigning the data points to $K$ clusters based on the current centroids and choosing the centroids based on the current assignment of the data points to clusters. It presents a scaling complexity of $O(KND)$, which corresponds to the dominating step where the $KN$ distances between all $D$-dimensional data points and all centroids are computed.

\vspace{0.5cm}

\subsection{Quantum \textit{k-means}}

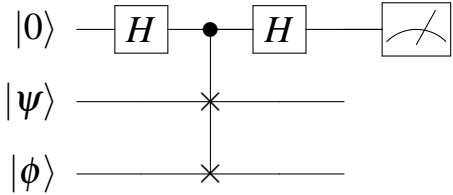
\begin{figure}[ht]
\hbox{\hspace{20em}
    \centering
    \leavevmode
    \Large 
        \Qcircuit @C=1em @R=0.5em @!R {
        \lstick{\ket{0}} & \gate{H} & \ctrl{1}  & \gate{H} \qw & \qw & \meter \qw  \\ 
        \lstick{\ket{\psi}} & \qw & \qswap & \qw &  \qw \\
        \lstick{\ket{\phi}} & \qw & \qswap \qwx & \qw &  \qw}}
    \vspace*{5mm}
    \caption{Quantum circuit corresponding to the \textit{SwapTest} routine.}
    \label{fig:SwapTest}
\end{figure}

In order to construct our digital quantum jet clustering algorithm based on quantum \textit{k-means}\cite{qkmeans}, one needs to start by identifying which of the classical algorithm's steps has the most potential to yield a quantum advantage. Here, this step corresponds to the distance calculation between the $N$ data points and the $K$ centroids. In order to compute the distances on a quantum circuit, the \textit{SwapTest}\cite{SwapTest} quantum sub-routine (see Figure \ref{fig:SwapTest}) is used\cite{Lloyd}. By measuring the overlap between two quantum states $\ket{\psi}$ and $\ket{\phi}$, $\braket{\psi|\phi}$, based on the measurement probability of the control qubit being in state $\ket{0}$, $P(\ket{0}) = \frac{1}{2} + \frac{1}{2}|\braket{\psi|\phi}|^2$, the \textit{SwapTest} routine is used to calculate the squared Euclidean distance $||\Vec{p}_i - \mu_k||^2$ between a particle's momentum vector $\Vec{p}_i$ and a given jet cluster's centroid $\mu_k$. To that purpose it performs the following steps:

\begin{enumerate}

    \item \textbf{State Preparation:} Prepare two quantum states,
        \begin{equation}
            \begin{split}
                & \ket{\psi} = \frac{1}{\sqrt{2}} \big( \ket{0,\Vec{p}_i} + \ket{1,\mu_k} \big) \\
                & \ket{\phi} = \frac{1}{\sqrt{Z}} \big( ||\Vec{p}_i||\ket{0} - ||\mu_k||\ket{1} \big), \qquad \text{\small with} \quad Z = ||\Vec{p}_i||^2 + ||\mu_k||^2
            \end{split}
            \label{Distance_states}
        \end{equation}
        
    \item \textbf{Find Overlap:} Compute overlap $|\braket{\psi|\phi}|^2$ through the \textit{SwapTest} sub-routine.
    
    \item \textbf{Compute Squared Euclidean Distance:} Get the desired squared Euclidean distance through the following equation,
        \begin{equation}
            \begin{split}
               ||\Vec{p}_i - \mu_k||^2 = 2Z|\braket{\psi|\phi}|^2
            \end{split}
            \label{Distance_calc}
        \end{equation}
        
\end{enumerate}
where the qubit registers $\ket{\Vec{p}_i}$ and $\ket{\mu_k}$ are prepared using Amplitude Encoding\cite{AmplitudeEnc1,AmplitudeEnc2}, or can be loaded directly from Quantum Random Access Memory (QRAM)\cite{QRAM}. Furthermore, the search for the closest jet centroid for each particle is also typically performed via the \textit{Grover Optimization} quantum sub-routine,\cite{GroverOpt} based on the original Grover search algorithm\cite{Grover}. However, given that this step does not affect the overall complexity of the algorithm and its jet clustering application proof of concept, we have chosen not to implement it. Consequently, the quantum algorithm's complexity is of the order of $O(KN\log D)$, resulting from the fact that only $\log D$ qubits are needed to encode both the particles' momentum vectors and jet centroids $\ket{\Vec{p}_i}$ and $\ket{\mu_k}$.

\vspace{0.5cm}

\subsection{Jet Clustering}

\subsubsection{Clustering Distance Metric}

The \textit{k-means} clustering algorithm usually aggregates points by taking into account how distant they are from each other. For this purpose, it makes use of the squared Euclidean distance, such that any points closer to each other tend to be clustered together. However, while performing this type of operation for jet clustering, it will tend to aggregate soft particles emitted in opposite directions despite the fact that they belong to opposing jets. This happens due to the small magnitude of their momenta. Furthermore, we know that if any two final-state particles $\Vec{p}_i$ and $\Vec{p}_j$ belong to the same jet, the angle $\theta_{ij}$ between the two particles will tend to be much smaller than to that of any other particle belonging to another jet. This is expected, since the large momenta of the produced quarks result in highly collimated jets such that $\theta(\Vec{p}_i,\Vec{p}_j) \ll \frac{\pi}{2}$ for any two particles $\Vec{p}_i$ and $\Vec{p}_j$ in the same jet. As such, we thus re-scale every particle's momenta to some multiple of the unit-sphere, such that we have a correspondence between the angle $\theta_{ij}$ between any two particles $\Vec{p}_i$ and $\Vec{p}_j$, and the distance $||\Vec{p}_i - \Vec{p}_j||$ between them. 

\subsubsection{\textit{A priori} Knowledge of $K$}

In order to successfully run the \textit{k-means} algorithm, it is known that one needs to provide the expected number $K$ of clusters to the algorithm. In the case of jet clustering, however, the number $K$ of jets produced is not known \textit{a priori}. Nevertheless, one does know the expected range of $K$ values as a function of the center-of-mass energy $\sqrt{s}$ and which particles are being collided. As such, we propose to run the algorithm a small number of times over the expected range for $K$, so that the most adequate number of jets can be inferred. We chose the value of $K$ which produces the highest quality clustering. For this work, we chose the \textit{Silhouette Index}\cite{Silhouette} as a figure of merit for clustering quality. By performing a quick complexity analysis of a given clustering's Silhouette calculation, however, we find that its computational cost is of the order of $O(N^2D)$, thus surpassing that of the algorithm itself. For this reason, a simplified \textit{Silhouette}  figure of merit is used, composed of the similarity measure $a(\Vec{p}_i)$, dissimilarity measure $b(\Vec{p}_i)$, and \textit{Silhouette} index $s(\Vec{p}_i)$ for each of the clustered particles:
\begin{equation}
    \begin{split}
        a(\Vec{p}_i) &= d(\Vec{p}_i,\mu_i)~, \\
        b(\Vec{p}_i) &= \min_{C_k \neq C_i} d(\Vec{p}_i,\mu_k)~, \\[3pt]
        s(\Vec{p}_i) &= \begin{cases}~
    \frac{b(\Vec{p}_i) - a(\Vec{p}_i)}{\max \big\{a(\Vec{p}_i),b(\Vec{p}_i) \big\}}, & \text{if $|C_i| > 1$}~,\\
    0, & \text{if $|C_i| = 1$}~. 
  \end{cases}
    \end{split}
    \label{Silhouette}
\end{equation}
where $C_i$ represents the jet cluster to which particle $\Vec{p}_i$ belongs. This way we have managed to reduce its computational cost to $O\big(N(K-1)\big)$, which scales slower than the overall algorithm. The overall clustering's \textit{Silhouette} is then obtained by computing  the mean of all $N$ particles' \textit{Silhouette} values:
\begin{equation}
    S_K = \frac{1}{N}\sum_{i}s(\Vec{p}_i)~. \label{totalS}
\end{equation}

\subsubsection{Performance Considerations}
The observables used in the clustering process are the three-momentum vectors of the particles. The dimensionality of the problem is thus constant with $D=3$. Since $D$ is constant, so is $\log D$ and this factor drops out in the calculation of the algorithmic complexity. The computational cost of the algorithm is thus simply $O(KN)$.

In what concerns the classical $k_t$ algorithm benchmark, we see that despite the possible naive $O(N^2)$ or even $O(N^3)$ implementations, it can be cleverly implemented in $O(N\log N)$ by exploiting some of its geometrical and minimum-finding aspects\cite{kt_scaling}. We now look at both the proposed algorithm and the $k_t$ benchmark's complexities, in order to understand how the two compare. The new proposed method becomes of interest only in the regime where the number of reconstructed jets $K \leq \log N$. 

Furthermore, it is important to notice that the $D=3$ dimensionality of the task at hand affects not only the scaling of the proposed quantum \textit{k-means} algorithm, but also its classical counterpart. Indeed, by dropping the $D$ factor, we obtain a complexity of $O(KN)$ for the classical \textit{k-means} algorithm. Since this is equivalent to that of its quantum analog, it can be said that the use of this quantum algorithm for real day-to-day jet clustering analysis becomes only relevant if one is able to exploit its dimensionality advantage of $\log D$ versus $D$ relative to the classical version. Although no such exploitation is proposed in this work, it should not be discarded, as interesting synergies with other stages of the jet clustering process are a strong possibility.

When measuring the algorithm's jet clustering efficiency, the ideal would be to compare it to the true jet regrouping for any given generated event, giving us information on the parenthood of each final-state particle and enabling us to know which particles should be clustered together. Unfortunately such Monte-Carlo truth is not available by design. Consequently, as mentioned above, we have chosen to measure the algorithm's performance against that of the classical $k_t$ algorithm. For a given clustering output, where the $N$ final-state particles have been sorted into $K$ jets, we compare both algorithm's clustering results on a particle-by-particle basis according to the following developed efficiency metric, $\epsilon$:
\begin{equation}
    \epsilon = \frac{\text{\# of particles grouped in the same way as $k_t$}}{\text{\# of particles in meaningful jets found by $k_t$}} \label{eff_metric}
\end{equation}
In order to identify the physically meaningful jets out of all the jets found by the $k_t$ algorithm, we apply a minimum transverse momentum $p_T$ jet cutoff, such that any given jet with transverse momentum lower than the set cutoff $p_T$ is discarded.

\section{Scaling of $K$ versus $\log N$}

We have used the \textit{PYTHIA}\cite{Pythia} Monte-Carlo (version 8.3) to generate the events on which the clustering should be performed (see Appendix for more details on event generation). To study the events' scaling of $K$ versus $\log N$, we have generated both $e^+e^- \rightarrow Z^0 \rightarrow q\bar{q}$ collision events at a center-of-mass energy of $\sqrt{s} = m_Z = 91.1876 \pm 0.0021$ $GeV/c^2$,\cite{PDG} as well as $pp$ collision events at center-of-mass energies of $\sqrt{s} = 7$ $TeV$ and $\sqrt{s} = 14$ $TeV$. We have also explored $pp$ collision events involving $t$-quarks given its high jet multiplicity. As such we have performed clustering on 1000 generated events of each kind, storing both the number of found meaningful jets $K$, as well as the corresponding event's logarithm of the number of final-state particles, $\log N$. 

\begin{figure}[ht]
    \centering
    \includegraphics[scale = 0.4]{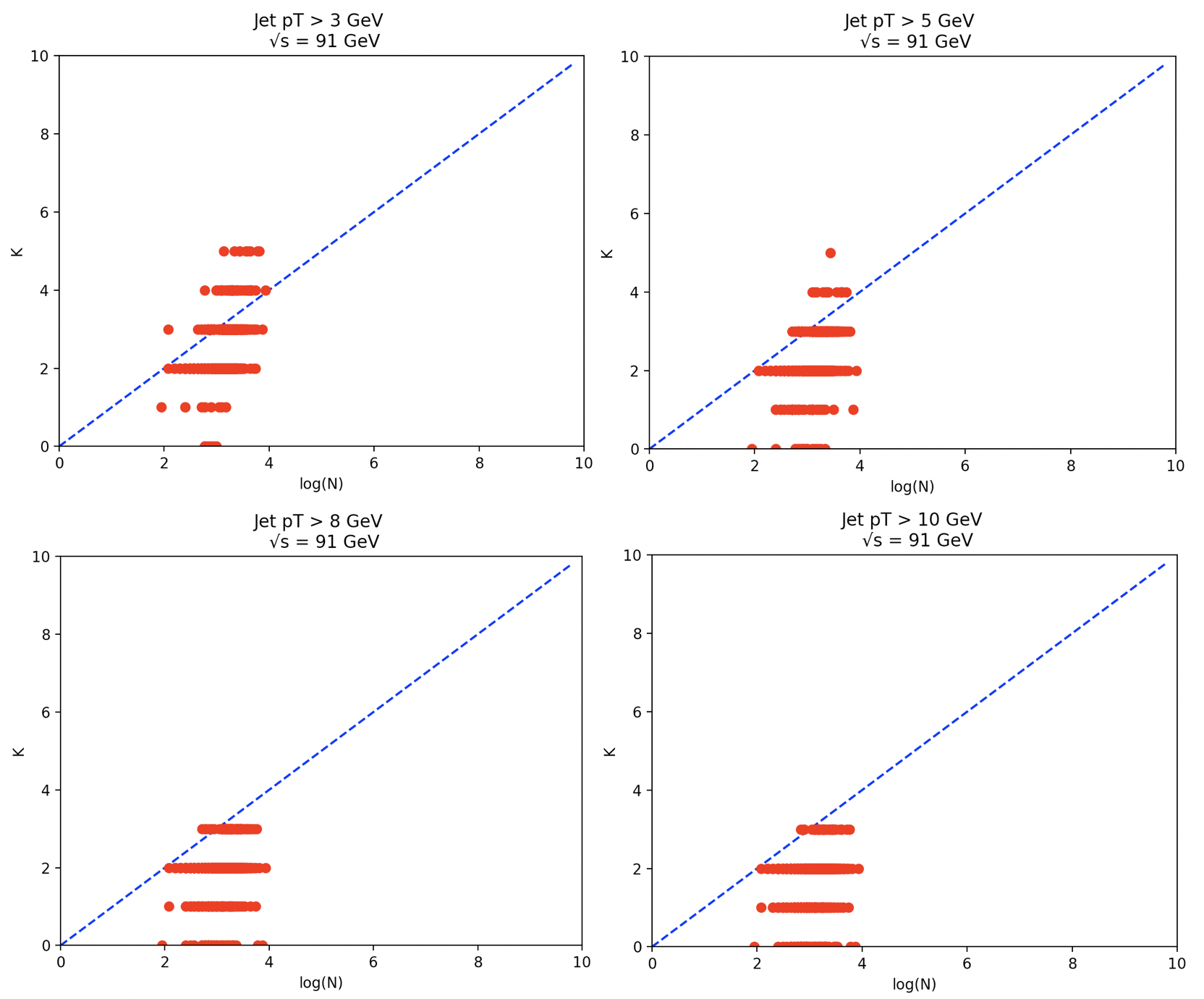}
    \caption{Plots of the number $K$ of found jets, against the logarithm of the number of final-state particles $\log N$ for four different jet $p_T$ cutoff scenarios in $e^+e^-$ collision generated events. Each red point represents a generated event, where $K$ jets have been found for the logarithm of $N$ particles.}
    \label{fig:scaling_ee}
\end{figure}

\begin{figure}[ht]
    \centering
    \includegraphics[scale = 0.36]{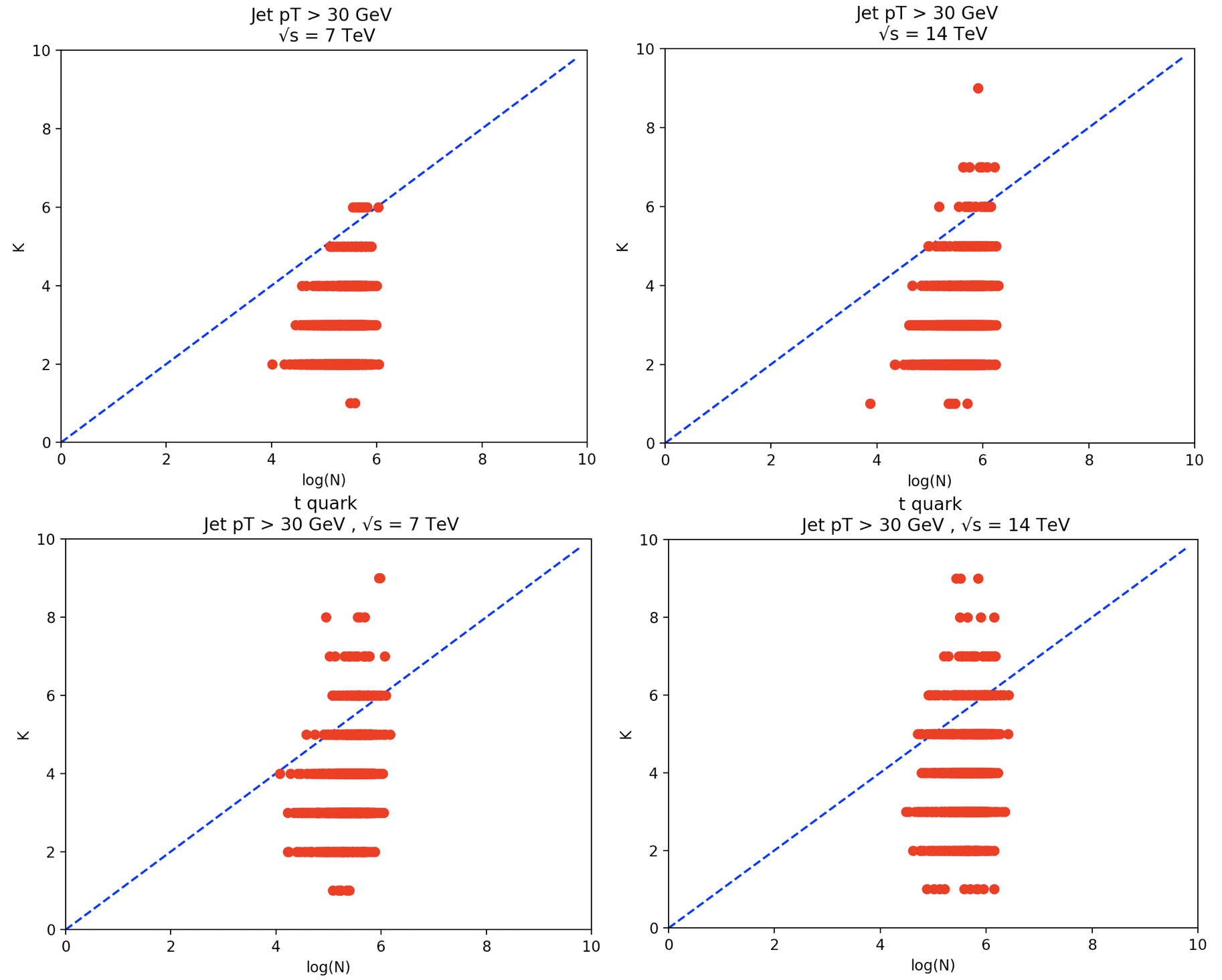}
    \caption{Plots of the number of jets found $K$, against the logarithm of the number of final-state particles $\log N$ for two different center-of-mass energies in $pp$ collision generated events. Each red point represents a generated event, where $K$ jets have been found for the logarithm of $N$ particles.}
    \label{fig:scaling_pp}
\end{figure}

We present the resulting plots in Figures \ref{fig:scaling_ee} and \ref{fig:scaling_pp}. The blue line dividing each of the plots in half represents the limit where $K = \log N$. As such, since we are hoping that $K \leq \log N$, ideally one would find the majority of events (red dots) below the blue line, indicating that indeed, the number of jets found is smaller than the logarithm of the number of final-state particles involved in the clustering process. From observation of the four plots of Figure \ref{fig:scaling_ee}, regardless of the chosen jet $p_T$ cutoff, it is clear that the majority of events (red points) already fall below the blue line, thus being in the regime where there is an advantage. Moreover, as expected, the higher we set the jet $p_T$ cutoff, the more events fall below the blue line, since we are allowing only higher $p_T$ jets to be accounted for, thus lowering the total number of jets found in each event. From observation of the plots of Figure \ref{fig:scaling_pp}, similarly to $e^+e^-$ collisions, the majority of events also fall below the blue line, thus finding themselves within the region of interest. As expected, both of the plots where the $t$ quarks are generated show higher jet multiplicities overall, nevertheless having most of the events falling below the line as well. It is also important to note that for a higher jet $p_T$ cutoff, higher $K$ levels would become less populated, moving down towards lower $K$ levels. This would lower the overall event multiplicity for all plots, thus resulting in an even higher portion of events falling within the region of $K \leq \log N$. We therefore conclude that a significant majority of events will yield a number of jets $K \leq \log N$, thus confirming our algorithmic complexity advantage.

\section{Clustering Performance}

Ideally, when applying the algorithm to the problem of jet clustering, one would want to explore its performance relative to both $e^+e^-$ and $pp$ collision events. However, given its unavoidably rudimentary implementation, the computational load of clustering $pp$ events is simply too high for the local CPU being used. As a result, we postpone these important test scenarios for future work, when a more robust version of the algorithm is likely to be developed.

\begin{figure}[ht]
    \centering
    \includegraphics[scale = 0.33]{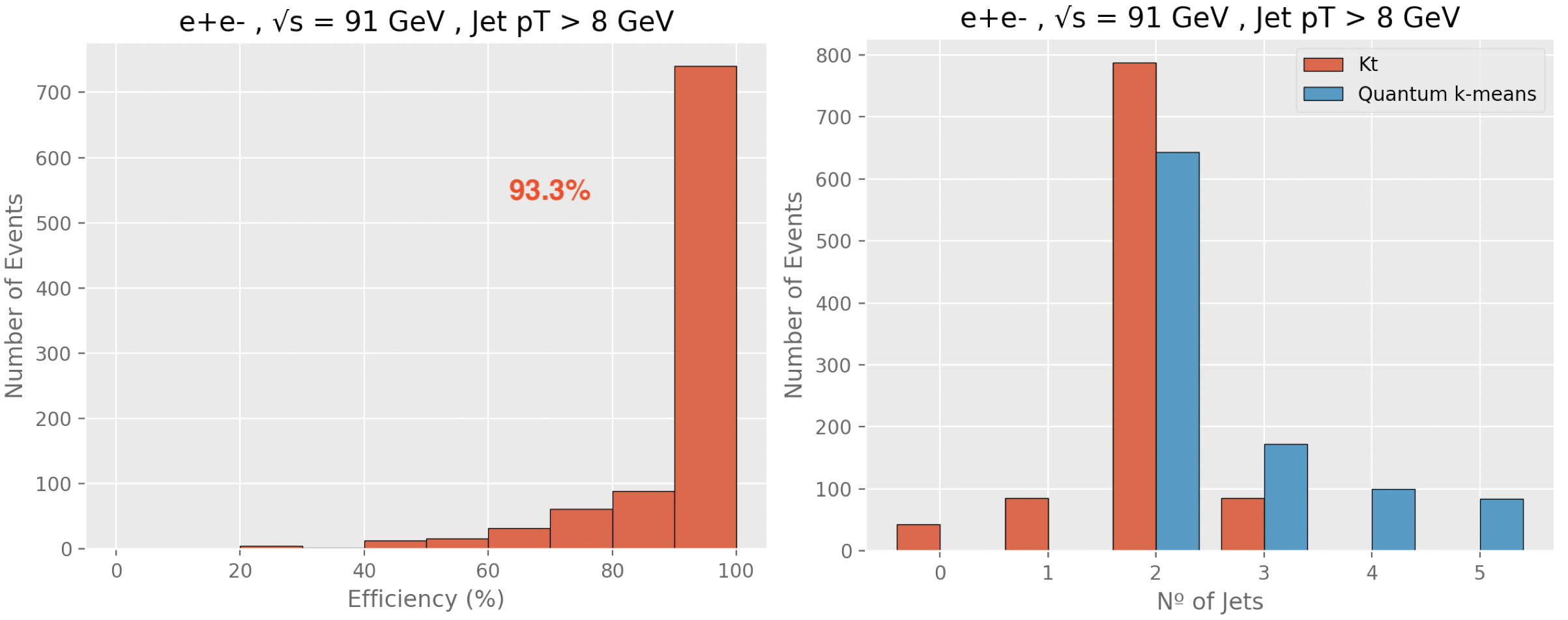}
    \caption{Histograms of the obtained efficiencies $\epsilon$ (on the left) and number of found meaningful jets (on the right) for a jet $p_T$ cutoff of $8$ $GeV$.}
    \label{fig:performance8}
\end{figure}

As such, we ran the proposed algorithm on the same 1000 events as above, studying both its clustering efficiency according to equation \eqref{eff_metric}, as well as its jet finding distribution, comparing it with the one obtained with the $k_t$ algorithm (see Appendix for more details on the $k_t$ clustering parameters and quantum \textit{k-means} implementation). The results for a jet $p_T$ cutoff of $8$ $GeV$ are presented in Figure \ref{fig:performance8}. From the left histogram, it can be seen that in the overwhelming majority of the clustered events, the quantum \textit{k-means} algorithm found same jet configurations as the $k_t$ benchmark, with a decreasing fraction of events for lower clustering efficiencies. The overall jet finding efficiency with respect to the $k_t$ algorithm is $\epsilon = 93.3\%$. Moreover, it can be seen from the right plot that the number of jets found by the $k_t$ algorithm is in the range from 0 to 3 while the proposed quantum \textit{k-means} algorithm ranges between 2 and 5. This is expected given that the high transverse momentum jet cutoff of $p_T = 8$ $GeV$ has been applied only to the $k_t$ algorithm, thus resulting in a lower number of overall found jets  . 

\begin{figure}[ht]
\hbox{\hspace{-2.5em}
    \centering
    \includegraphics[scale = 0.35]{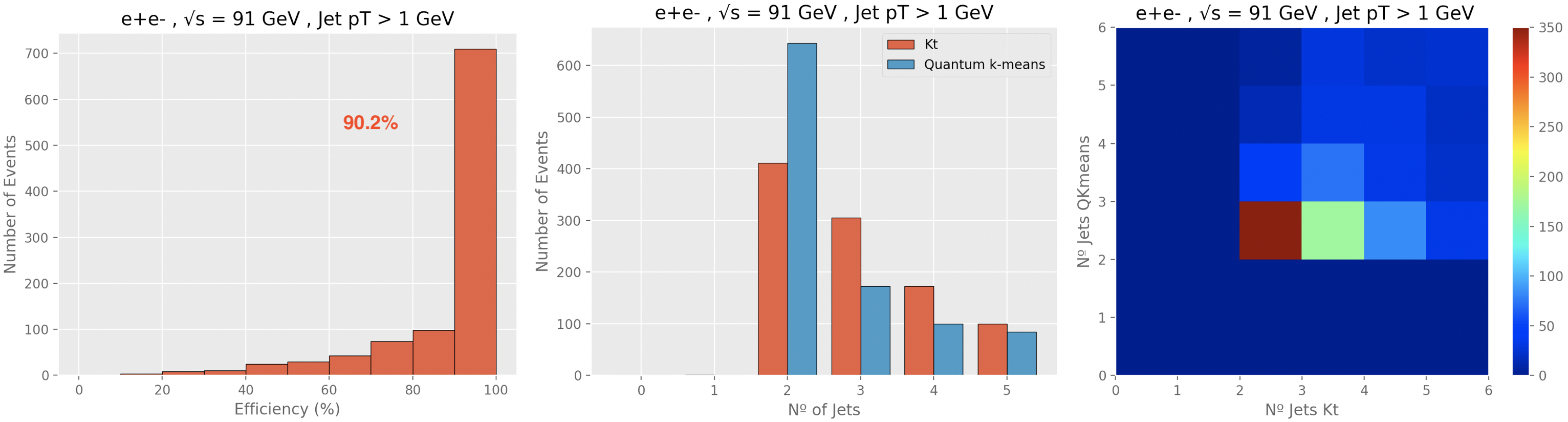}}
    \caption{Histograms of the obtained efficiencies $\epsilon$ (on the left), number of found meaningful jets $K$ (middle), and heatmap of the number of jets found by the proposed algorithm against those found by the $k_t$ benchmark for a jet $p_T$ cutoff of $1$ $GeV$.}
    \label{fig:performance1}
\end{figure}

To better understand the relation between the two algorithms, the applied jet $p_T$ cutoff was lowered to $p_T = 1$ $GeV$, with the purpose of artificially imposing a near zero barrier to the number of meaningful jets found by the $k_t$ algorithm. The resulting plots can be found in Figure \ref{fig:performance1}. As before, a very high efficiency of $\epsilon = 90.2\%$ has been obtained, showing that even for a significantly larger number of jets (see right plot) found by the $k_t$, the clustering efficiency has remained nearly as high. Regarding the distribution of the number of jets found, we can now see from both the righthand histogram and the heatmap plot, that there is a strong correlation between the number of jets found by both algorithms.

\section{Conclusions}

In this work, we have introduced the first digital quantum algorithm for jet clustering in high-energy physics, representing an alternative to the existing quantum annealing state-of-the-art\cite{Harrow,Pires}, and opening the doors to harnessing the power of future digital quantum processors to address this increasingly complex problem. Our algorithm yields jet reconstruction efficiencies of the order of $93\%$. Furthermore, we have found a clear correlation with the results of the $k_t$ algorithm in the number of jets found, further validating our quantum algorithm.

It is interesting to note the differences between both algorithms in number of events for each $K$ (see Figure \ref{fig:performance1}, right plot), where a larger number of events with smaller $K$ (mostly $K=2$) has been found by the quantum algorithm. This can be explained by the choice of the \textit{Silhouette} figure of merit: the proposed algorithm showed a tendency to systematically output a smaller $K$. This can be improved by a suitable choice of a figure of merit adequate to the nature of the task at hand. It is also important to realize that given the efficiency metric used, it is possible that for a portion of the events, the quantum algorithm performs better than the $k_t$ algorithm benchmark, which lowers its efficiency versus the latter, when in fact it should be increasing. Although unfortunate, this effect cannot be quantified.

On the other hand, despite being currently on equal footing relative to its classical counterpart, the quantum algorithm has been shown to benefit from the fact that $K < \log N$ for the majority of events, thus scaling better than its $k_t$ algorithm benchmark. It is nevertheless clear that there is great potential to improve the jet clustering process by exploiting the dimensionality advantage of $\log D$ versus $D$ of our quantum algorithm. 

Furthermore, our digital quantum algorithm motivates investigating other challenges such as outlier particles (particle remnants that do not belong to any jet) and pileup. The results of these investigations will be reported in future publications.

%TC:ignore

\bibliography{sample}

\section*{Acknowledgements}

The authors would like to thank Akshat Kumar, Duarte Magano, Adam Glos and Jesse Thaler for their valuable feedback. YO thanks the support from Funda\c{c}\~{a}o para a Ci\^{e}ncia e a Tecnologia (Portugal), namely through project UIDB/50008/2020, as well as from projects TheBlinQC and QuantHEP supported by the EU H2020 QuantERA ERA-NET Cofund in Quantum Technologies and by FCT (QuantERA/0001/2017 and QuantERA/0001/2019, respectively), and from the EU H2020 Quantum Flagship project QMiCS (820505). JS would like to thank the support of FCT under contracts CERN/FIS-COM/0036/2019 and UIDB/04540/2020.

\section*{Appendix}

\subsection*{\textit{PYTHIA} Event Generation}

For the event generation, as already mentioned, we have used \textit{PYTHIA}\cite{Pythia}, where different event generation settings have used for $e^+e^-$ and $pp$ collisions. Starting by $e^+e^-$ collision events, we have studied events of the type $e^+e^- \rightarrow Z^0 \rightarrow q\bar{q}$, such that all $Z^0$ decays have been switched off with only those to quarks having been manually switched on. Implicit by the center-of-mass energy used of $\sqrt{s} = m_Z$, only $q\bar{q} \in \{u\bar{u},d\bar{d},c\bar{c},s\bar{s},b\bar{b}\}$ decays have been allowed, since the $t$ quark is too massive for the center-of-mass energy used here ($m_t \ll \sqrt{s} = m_Z$).

For $pp$ collisions, all hard QCD processes were switched on through the flag \textit{HardQCD:all = on}. Moreover, a minimum invariant $p_T$ threshold of $200$ $GeV$ was set, through the parameter \textit{PhaseSpace:pTHatMin = 200.} For the $t$ quark processes studied, all hard QCD processes were turned off, while $t$ quark processes were switched on through the flag \textit{Top:all = on}. No further constraints or settings were imposed besides the collision's center-of-mass energy definition of $\sqrt{s} \in \{7,14\}$ $TeV$, thus resembling LHC conditions.

\subsection*{$k_t$ Clustering Parameters}

The $k_t$ clustering algorithm has been implemented and used through the \textit{FastJet} software package\cite{Fastjet,kt_scaling}. By using the Jet Definition \textit{jet\_def(kt\_algorithm, R)}, the $k_t$ clustering algorithm has been selected and chosen to run with an $R$ parameter of $R = 0.8$. It is of interest to note that for values of $R \in [0.4,1]$, the clustering efficiency of the quantum algorithm has remained constant. The $k_t$ algorithm received as input \textit{PYTHIA} generated events' final-state particles' momentum coordinates plus their corresponding energy. When outputting the jets found, jet $p_T$ cutoffs of $p_T \in [1,15]$ $GeV$ were applied by selecting only those jets with $p_T$ higher than the defined cutoff, again without much variation on the resulting efficiencies.

\subsection*{Quantum \textit{k-means} Implementation}

Taking into account the current cloud-available Quantum Processing Unit (QPU) hardware constraints, we have chosen to use the local CPU in order to simulate quantum QPU behaviour with the help of Qiskit's appropriate software packages, \textit{Terra} and \textit{Aer}. The proposed quantum \textit{k-means} algorithm should be seen as a hybrid algorithm, meaning that only a portion of it is performed on a digital quantum computer, with the rest of it running on a classical machine. In this particular case, only the distance computation between particles and centroids has been implemented to run under the Qiskit's quantum QPU behavior simulations package while the remainder of the algorithm has been coded from scratch to be classical. 

Given the inherently large nature of jet clustering problems in HEP and the fact that IBM's \textit{qasm\_simulator} feature (which allows QPU behavior simulation on the local CPU) is prepared for the simulation of a maximum of 32 qubits, we have chosen to implement one single 5-qubit \textit{SwapTest} circuit, running $\sim10000$ times for it to calculate every distance needed in each iteration. We are therefore implicitly trading off computational speed for feasibility, allowing us to proceed with the implementation of the algorithm. Furthermore, it would be ineffective to try to simulate any more than $\sim 24$ qubits on the classical machine that is being used (15-inch, 2017 MacBook Pro, with a 3.1 GHz Intel Quadcore i7 CPU, and 16GB of RAM).

Even though it is trivial to initialize the control qubit, given that it corresponds simply to a one qubit register $\ket{0}$, the same cannot be said of the qubit registers $\ket{\psi}$ and $\ket{\phi}$. Indeed, given their definitions in \eqref{Distance_states}, one has to put a bit of effort into their preparation. Fortunately, Qiskit allows for a very straightforward and useful initial state definition, simply by feeding it each of the states' amplitudes. Let us start by the simpler case of the quantum state $\ket{\phi}$. Given its definition, we can write:

\begin{equation}
\begin{split}
    \ket{\phi} &= \frac{1}{\sqrt{Z}} \big( ||\Vec{p}_i||\ket{0} - ||\mu_k||\ket{1} \big) \\
    &= \frac{||\Vec{p}_i||}{\sqrt{Z}}\begin{bmatrix} 1 \\ 0 \end{bmatrix} - \frac{||\mu_k||}{\sqrt{Z}}\begin{bmatrix} 0 \\ 1 \end{bmatrix} \\
    &= \frac{1}{\sqrt{Z}}\begin{bmatrix} ||\Vec{p}_i|| \\ -||\mu_k|| \end{bmatrix}
\end{split}
\label{initialize_phi}
\end{equation}

On the other hand, assuming Amplitude Encoding\cite{AmplitudeEnc1,AmplitudeEnc2}, when it comes to the quantum state $\ket{\psi}$, one needs first to adequately express the quantum states corresponding to particle $\Vec{p}_i$ and jet centroid $\mu_k$, $\ket{\Vec{p}_i}$ and $\ket{\mu_k}$. We thus write for an arbitrary data point $a$:

\begin{equation}
\begin{split}
    \ket{a} &= \frac{1}{|a|}\sum_{i=1}^D a_i\ket{i} \\
    &= \frac{a_x}{|a|}\ket{00} + \frac{a_y}{|a|}\ket{01} + \frac{a_z}{|a|}\ket{10} \\
    &= \frac{1}{|a|}\begin{bmatrix} a_x \\ a_y \\ a_z \\ 0 \end{bmatrix}
\end{split}
\end{equation}

As such, for the quantum state $\ket{\psi}$, we can now write:

\begin{equation}
\begin{split}
    \ket{\psi} &= \frac{1}{\sqrt{2}} \big( \ket{0,\Vec{p}_i} + \ket{1,\mu_k} \big) \\
    &= \frac{1}{\sqrt{2}} \Bigg( \begin{bmatrix} 1 \\ 0 \end{bmatrix} \otimes \frac{1}{||\Vec{p}_i||} \begin{bmatrix} p_x \\ p_y \\ p_z \\ 0 \end{bmatrix} + \begin{bmatrix} 0 \\ 1 \end{bmatrix} \otimes \frac{1}{||\mu_k||}\begin{bmatrix} \mu^k_x \\ \mu^k_y \\ \mu^k_z \\ 0 \end{bmatrix} \Bigg) \\
    &= \frac{1}{\sqrt{2}} \begin{bmatrix} \frac{p_x}{||\Vec{p}_i||} & \frac{p_y}{||\Vec{p}_i||} & \frac{p_z}{||\Vec{p}_i||} & 0 & \frac{\mu^k_x}{||\mu_k||} & \frac{\mu^k_y}{||\mu_k||} & \frac{\mu^k_z}{||\mu_k||} & 0 \end{bmatrix}^T
\end{split}
\label{initialize_psi}
\end{equation}

Now, through equations \eqref{initialize_phi} and \eqref{initialize_psi}, one can simply take the respective vectors' elements and feed them to the Qiskit \textit{initialize()} function, thus successfully generating the desired quantum states $\ket{\psi}$ and $\ket{\phi}$.

% \section*{Author contributions statement}

% Must include all authors, identified by initials, for example:
% A.A. conceived the experiment(s),  A.A. and B.A. conducted the experiment(s), C.A. and D.A. analysed the results.  All authors reviewed the manuscript. 

% \section*{Competing Interests}

% The corresponding author is responsible for submitting a \href{http://www.nature.com/srep/policies/index.html#competing}{competing interests statement} on behalf of all authors of the paper. This statement must be included in the submitted article file.

% \section*{Additional information}

%TC:endignore

\end{document}